\documentclass[sigconf, nonacm]{acmart}
\usepackage{pifont}
\usepackage{xcolor}
\usepackage{hyperref}
\usepackage{algorithm}
\usepackage{algpseudocode}
\usepackage{stfloats}
\usepackage[most]{tcolorbox} 
\usepackage{subcaption}
\usepackage{makecell}
\usepackage{ulem}

\newcommand{\mountdb}{MountDB} 
\newcommand{\rocksdb}{RocksDB} 
\newcommand{\bourbon}{Bourbon} 
\newcommand{\wisckey}{WiscKey} 
\newcommand{\alexol}{ALEX$+$} 
\newcommand{\skiplist}{SkipList}
\newcommand{\pgm}{PGM} 
\newcommand{\tridentkv}{TridentKV} 
\newcommand{\doblix}{DobLIX}
\newcommand{\pgmfence}{PGM (Fence)}

\newtcolorbox{takeawaybox}{
  enhanced,                 
  colback=gray!12,
  colframe=gray!12,
  boxrule=0pt,
  arc=0pt,
  left=6pt,
  right=6pt,
  top=6pt,
  bottom=6pt,
  borderline west={2pt}{0pt}{black} 
}

\usepackage{enumitem}
\def\shorten{\looseness=-1}

\begin{document}
\title{A Pragmatic Approach to Learned Indexing in \rocksdb{}: \\ Targeted Optimizations with Minimal System Modification}
\author{Shubham Vashisth}
\affiliation{%
  \institution{McGill University}
  \city{Montréal}
  \state{Canada}
}
\email{shubham.vashisth@mail.mcgill.ca}

\author{Olivier Michaud}
\affiliation{%
  \institution{McGill University}
  \city{Montréal}
  \state{Canada}
}
\email{olivier.michaud2@mail.mcgill.ca}

\author{Bettina Kemme}
\affiliation{%
  \institution{McGill University}
  \city{Montréal}
  \state{Canada}
}
\email{bettina.kemme@mcgill.ca}

\author{Oana Balmau}
\affiliation{%
  \institution{McGill University}
  \city{Montréal}
  \state{Canada}
}
\email{oana.balmau@mcgill.ca}

\begin{abstract}

Learned indexes emerged as a promising alternative to classic index structures, offering higher throughput and reduced memory footprint by approximating the cumulative key distribution function through lightweight models. Despite these advantages, learned indexes have seen limited adoption in production systems. One probable reason is that learned indexes that handle concurrent updates and persistence as effectively as, e.g., the B$^+$-Tree, do not yet exist. Moreover, many research prototypes introduce complex and difficult-to-maintain solutions. In this paper, we take a pragmatic approach to these issues and explore whether we can integrate off-the-shelf learned indexes into a production database system with minimal redesign of the storage system. We use \rocksdb{}, a widely used key-value store with the popular log-structured merge (LSM) architecture, as a case study. \rocksdb{} has a clear separation of main memory based Memtables that ingest writes, and read-only sorted files on disk. This allows us to deploy specialized indexes at each level without the need for a "can-do-all" solution. Our investigation shows that it is not enough to simply use an existing index out-of-the-box under write-heavy workloads, where the frequent replacement of Memtables goes against the learning nature of the index, which needs warm-up time to work well. 
We address this with an effective reuse mechanism that preserves structural knowledge across Memtable instances. At the disk-level, we replace \rocksdb{}'s index with a learned index without any changes to the remainder of the storage layer or read path. For that, we adapt an effective in-memory read-only index to be block-aware, enabling worst-case single-I/O lookups.
We incorporate these findings in an extension of \rocksdb{} we call \mountdb{}. Our evaluation on large-scale workloads with diverse data distributions and access patterns demonstrates that  \mountdb{} achieves up to $1.5\times$ faster writes, and $2.1\times$ faster reads 
compared to state-of-the-art systems, showing that established learned indexes can be incorporated into production systems with negligible overhead and promising results.\shorten{}

\end{abstract}

\maketitle
\pagestyle{empty}



\begin{figure}[t]
  \centering
  \includegraphics[width=1.01\linewidth]{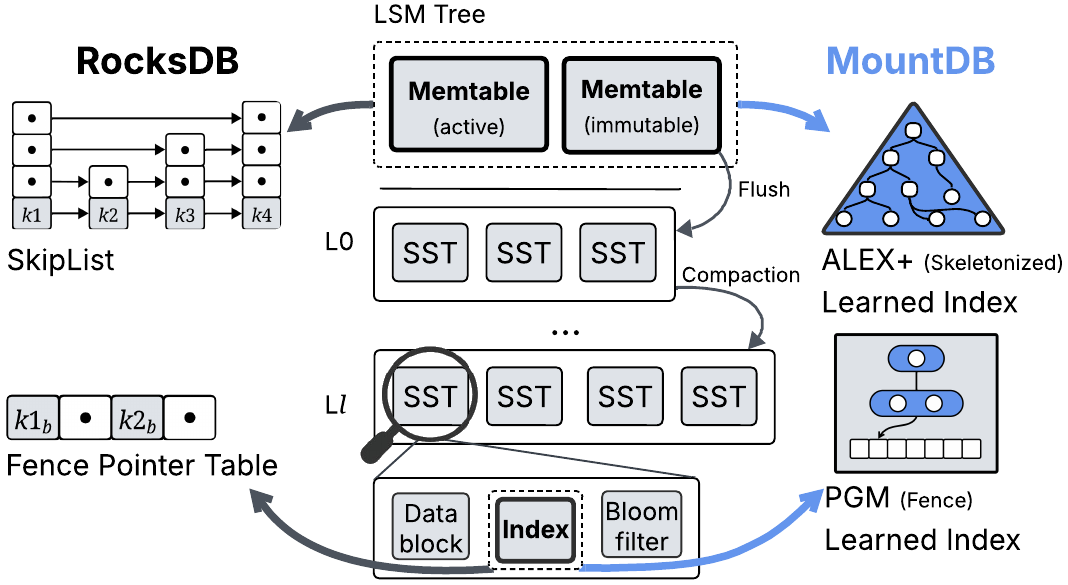}
  \caption{Indexing in LSMs. \rocksdb{} uses classic index structures while \mountdb{} adopts optimized learned indexes.\shorten{}}
  \vspace{-3mm}
  \label{fig:lsm-mountdb}
\end{figure}

\section{Introduction}
\label{sec:introduction}
Since the seminal paper on learned indexes appeared in 2018 \cite{learned-index}, there has been a significant body of research on developing learned index structures. This work claims impressive improvements over classic indexes \cite{fitting_tree, pgm, xindex, learned_indexes_survey, art}. Still, nearly a decade later, learned indexes are not yet used in production systems, and the question arises whether they are missing something fundamental that makes them unsuitable for today's data storage systems. Crucially, it is unclear whether the impressive standalone performance benefits improve end-to-end performance when integrated into a full-fledged database system. There is also the question whether integration is too complicated or too daring, given that much prior work proposes sophisticated policies for learned index integration~\cite{bourbon}, or building learned databases from the ground up~\cite{kraska2021sagedb}. This paper explores these questions, with Log-Structured Merge (LSM) key-value stores as a case-study.

We take a pragmatic approach by analyzing the feasibility of learned indexing in a production environment. More precisely, we look at the integration of learned indexing into RocksDB \cite{rocksdb}, a popular key-value store (e.g., used at Meta, Uber, Airbnb, Netflix, Nutanix).  RocksDB is a Log-Structured Merge (LSM) based datastore. LSM trees are also the backbone of many other storage engines such as LevelDB \cite{leveldb}, and Cassandra \cite{cassandra}. LSMs follow a two-component architecture (see Figure \ref{fig:lsm-mountdb}) that balances write efficiency with query performance. To handle writes, LSMs maintain small sorted in-memory Memtables that ingest updates. Memtables are flushed as immutable, read-only Sorted String Table (SST) files to disk. Background compaction then merges SSTs to maintain key order and reclaim space. Reads first consult the Memtable and then search across the SSTs. To ensure acceptable read latencies, both Memtables and SSTs have their own indexes. 

This two-component design of LSMs provides a compelling opportunity to integrate different specialized learned indexes across the two components, choosing for each of them an index whose characteristics fit its local requirements. For the Memtable, we need an updatable concurrent main-memory index, while we can use read-efficient static indexes for the persistent, read-only SSTs. Thus, while a holistic general-purpose learned index that can compete with B+-trees both in-memory and on disk might not yet exist, current solutions might suffice for LSMs. 

Prior work has integrated learned indexes into the persistence layer (SSTs) of LSMs, in systems such as \bourbon{}~\cite{bourbon}, Doblix~\cite{doblix}, TridentKV~\cite{tridentkv}, and LeaderKV~\cite{leaderkv}. While these solutions are promising from a performance point of view, their integration efforts are significant: they either apply complex rules to decide on index creation and management~\cite{bourbon,doblix}, or they redesign the LSM storage structure to accommodate the indexes~\cite{leaderkv,tridentkv,doblix}. 

We believe that neither approach is appealing for production systems, as a modification to the core of the storage infrastructure would be risky to adopt. We take the opposite approach. Our guiding principle is to integrate learned indexes  with as few changes as possible to the existing software stack. To find the right indexes and deploy them efficiently, we analyze the operational constraints of LSMs in-depth. We find two key challenges: one at the memory component, and one at the disk component. First, in LSMs, Memtables are fairly small. This leads to the creation and flushing of many Memtables, resetting the index -- a situation that places classic updatable learned indexes at a disadvantage. Second, as noted in prior work~\cite{pgm_relaxed_error_bound, aulid}, special care must be taken to align learned indexes to block-level I/O when using learned indexes for on-disk data. It is a challenge to do so without changing the entire structure of the SSTs (e.g., as done by Doblix~\cite{doblix} and TridentKV~\cite{tridentkv}). Finally, as each Memtable and SST file needs an index, their creation must have a low overhead and not impact read/write performance. There is potential to piggyback on flushing and compaction operations, creating the indexes off the critical path.

The result is \mountdb{}, an LSM key–value store built as an extension of \rocksdb{}. \mountdb{} is the first LSM-based system to employ learned indexes across both memory and disk levels, in conjunction with targeted optimizations specifically designed to address LSM-inherent design constraints. Importantly, \mountdb{} introduces minimal modifications to the existing architecture as seen in Figure~\ref{fig:lsm-mountdb} (right), where only the index components of \rocksdb{} are replaced, thereby preserving the proven advantages of the production system while enabling performance benefits of learned indexing. \shorten{}

For its write-heavy Memtables, \mountdb{} uses \alexol{} \cite{gre}, an updatable learned index that efficiently supports in-place modifications and concurrent inserts. A major issue with \alexol{} and many other learned indexes, is that they have been designed for steady-state workloads and perform best when a large number of records already exist at their creation time, for initialization. However, in LSMs, whenever a new empty Memtable is created, a fresh index accompanies it. To alleviate these cold-start inefficiencies of \alexol{}, \mountdb{} introduces a skeletonization optimization that preserves the index structure across Memtable flushes.\shorten{}

At the disk level, \mountdb{} adopts the \pgm{} index \cite{pgm}, which offers bounded and fast search capabilities for immutable data. As \pgm{} is a main-memory index that predicts the position of a record in an array we adjust it to instead predict the block a record resides in. At the same time, we ensure that the index itself remains small and we perform bulk I/O~\cite{pgm_relaxed_error_bound} in order to accommodate wrong predictions.
\mountdb{} takes the overhead of creating the indexes off the critical path of user requests. 
Together, these choices allow \mountdb{} to integrate learned indexes across the LSM hierarchy with throughput and latency improvements across-the-board without introducing significant complexity to the workflow. 

In summary, the key contributions of this paper are as follows:\shorten{}

\begin{itemize}[leftmargin=*, nosep]

\item We present a study showing when state-of-the-art learned indexes might fail when used out of the box in LSMs.\shorten{}

\item We design and implement \mountdb{}, an LSM system integrating learned indexes into \rocksdb{} with minimal, targeted modifications. 

\item We provide an extensive experimental evaluation of \mountdb{} across six real-world datasets, examining concurrency, varied value sizes, and ablation studies of key design choices. Compared to state-of-the-art learned index solutions in LSMs - \doblix{} \cite{doblix}, and \tridentkv{} \cite{tridentkv}, \mountdb{} achieves up to $\sim$1.5$\times$ faster writes and $\sim2.1\times$ faster reads, demonstrating that learned indexes can be both practical and efficient in modern LSMs.
\end{itemize}

\section{Background and Related Work}
\label{sec:background}

\subsection{Log Structure Merge (LSM) Trees}


The high-level structure of a Log-Structured Merge Tree \cite{lsm} is presented in Figure~\ref{fig:lsm-mountdb} (left). LSMs were introduced to efficiently handle write-intensive workloads by exploiting the advantages of both memory and disk-based structures. LSMs employ an append-only approach: updates create new record versions that are appended to an in-memory structure called a Memtable (typically between 64-128 MB). When the active Memtable is full, it becomes immutable, and a new, empty Memtable is created. The default Memtable in \rocksdb{} maintains a concurrent \skiplist{} \cite{skiplist} that supports logarithmic lookup of records. Its probabilistic, linked-list-like structure allows for straightforward implementation of non-blocking concurrent operations. Its average-case look-up efficiency, compact meta-information, and simplicity make it a robust and practical index choice for the transient, in-memory component of an LSM.\shorten{}

The disk-based component consists of multiple levels of Sorted String Tables (SST).  Each of them contains a set of records sorted by key. An immutable Memtable is flushed to disk into an SST at the highest level (L0).
As an upper level fills up, a background compaction process merges SSTs from this level with SSTs in the next lower level, in a merge-sort fashion, removing stale records. This hierarchically organizes data, with newer / hot data in smaller, upper levels and older / cold data in larger, lower levels. \rocksdb{} employs a Fence pointer table as the on-disk index, storing one entry per data block in each SST. Each entry consists of the block’s first key and a pointer to its location, leading to the index growing proportionally with the number of blocks. This structure is agnostic to the underlying distribution of keys. Additionally, \rocksdb{} uses Bloom filters to avoid accessing files that do not contain the desired key \cite{bloom_filter}.\shorten{}

A lookup operation first queries the active Memtable, followed by any immutable Memtables, before proceeding to on-disk SST files starting with level L0. First, the Bloom filter is checked for existence, and then the Fence pointer table is used to locate the data block containing the key-value pair. Within the data block, \rocksdb{} performs a binary search for the record. Given the default block size of around 4 KB, binary search is efficient as these blocks usually contain relatively few records. If the key is not found at level L0, the search continues to deeper levels. 
This design enables high write throughput while maintaining reasonable read performance. Consequently, LSM-based key-value stores have seen widespread adoption. \shorten{}



\subsection{Learned Indexes}
Learned indexes reinterpret indexing as a machine learning regression problem: given a set of keys, the index learns the cumulative distribution function of the key distribution and uses it to typically predict the position of a key in a sorted array. A final last-mile search corrects residual errors \cite{learned-index}. This design shifts complexity from a conventional tree traversal to model inference and local refinement, yielding smaller indexes and faster lookups. A foundational learned index design is the Recursive Model Index (RMI) \cite{learned-index}, which organizes a hierarchy of models. The root model predicts which sub-model to use, and this process repeats until a leaf model outputs a position. \shorten{}

\vspace{1mm}
\noindent\textbf{Read-optimized learned indexes} have static, read-optimized designs, such as the \pgm{} index \cite{pgm}. They target immutable datasets. \pgm{} uses piecewise linear models (PLA) to approximate the CDF with a strict error bound $\varepsilon$. The construction minimizes the number of models (segments) needed to satisfy this error, achieving both high compression and deterministic performance: a query touches only a bounded window of $\pm \varepsilon$ keys. The recursive structure of \pgm{} extends this compression hierarchically, producing one of the most space-efficient indexes with provable guarantees \cite{old_pgm, pgm, gre}. 

\vspace{1mm}
\noindent\textbf{Updatable learned indexes} 
extend the RMI structure to support efficient insertions and updates. ALEX \cite{alex} organizes data in gapped arrays, sorted arrays with strategically placed gaps that locally absorb insertions, minimizing costly full-array shifts. Insertions leverage model-based placement, positioning keys near their predicted locations to maintain spatial locality and reduce future prediction errors. To sustain performance under updates, ALEX employs lightweight cost models that trigger adaptive structural adjustments when nodes become full. 
Model accuracy is preserved through selective retraining or scaling during expansions, guided by cost deviations rather than periodic reorganization. This design allowed ALEX to support updates efficiently without compromising its read performance. Prior work shows that its best performance is achieved when the model is created by inserting a bulk of data in one shot~\cite{gre}. 
\alexol{} \cite{gre} extends ALEX by introducing concurrency. It combines fine-grained synchronization with localized updates in gapped arrays, enabling multiple threads to read and write concurrently with minimal blocking. This structure employs optimistic locking at the leaf level, with one lock per data node, and uses per-node shared-exclusive locks for internal nodes \cite{apex}, allowing safe and efficient parallel operations without undermining the original performance benefits of ALEX.

\vspace{1mm}
\noindent\textbf{Persistent learned indexes} are adaptations of learned indexes to disk storage. Naive adaptation of in-memory indexes to disk failed to outperform a disk-based B+Tree~\cite{aulid_prelimiary}. The performance gap was driven less by model accuracy and more by storage-layer overheads, including I/O-heavy scans, costly structural modifications, and the maintenance of internal data structures, such as gapped arrays and insertion buffers that were poorly aligned with block-based storage~\cite{alex}.
To address these challenges, AULID~\cite{aulid} was introduced as a disk-native learned index that combined model-based inner nodes with B+Tree-inspired leaf nodes. This hybrid design reduced tree height to lower I/O while preserving the efficiency of B+Trees for leaf operations such as updates, splits, and sequential scans \cite{aulid_prelimiary}. 

A complementary line of work has focused on systematically transforming in-memory learned indexes into disk-efficient counterparts. A recent study~\cite{pgm_relaxed_error_bound} proposes a set of design principles that guide this transformation, emphasizing techniques such as aligning model error bounds with block boundaries to optimize the last-mile search, prefetching candidate blocks to reduce I/O, compressing model parameters, and reliance on hybrid buffering to handle updates efficiently.
These advances highlight that persistent learned indexes may be feasible, either through disk-native designs like AULID or careful transformations of existing structures.


\vspace{1mm}
\noindent\textbf{Learned indexes in LSMs} were first explored in \bourbon{} \cite{bourbon}, which extends \wisckey{} \cite{wisckey}, an LSM system that reduces the high write costs of compactions by performing the separation of keys and values. \bourbon{} only uses learned indexes in the LSM read-only disk component, and for selected SSTs. \bourbon{}  assumes that index creation is costly and might not be worth for short-lived SSTs. As such, it employs a cost–benefit analyzer that observes the read workload to decide when an index is worthwhile constructing. It then uses greedy piecewise linear regression (Greedy-PLR) \cite{greedy_plr} to approximate key distributions and speed up lookups.
Its indexing method offers less predictable performance than later approaches, such as the \pgm{} index \cite{old_pgm}, and sometimes even increases tail latency \cite{tone,doblix}. Finally, its cost-benefit analyzer adds considerable complexity.\shorten{}

Subsequent research has built upon \bourbon{}'s foundation, each addressing its limitations. TridentKV \cite{tridentkv} introduces an adaptive training strategy to avoid write-path bottlenecks and redesigns the storage architecture by introducing a partitioning scheme to handle deletions more efficiently. LeaderKV \cite{leaderkv} also takes a structural approach, co-designing a new key–value storage layout alongside a hybrid learned index with redirect mechanisms to contain misprediction costs. DobLIX \cite{doblix} introduces a self-tuning framework that leverages reinforcement learning to dynamically balance index accuracy against the I/O cost. While effective, these systems deploy complex modifications to the LSM's disk component.
Moreover, they all remain focused exclusively to the disk, leaving the in-memory indexing of LSMs unexplored. \shorten{}

\section{Challenges with Existing Learned Indexes}
\label{sec:design_tensions}





As discussed above, recent advances in learned index design have produced specialized structures for different workload characteristics \cite{gre, loft, pgm}. A natural approach is therefore to directly integrate state-of-the-art learned indexes into each LSM component: an updatable and concurrent structure, such as \alexol{}~\cite{gre} for Memtables, and a read-optimized one such as \pgm{}~\cite{pgm} for SSTs.\shorten{}

However,  
when working within a full-fledged system, a simple plug-and-play approach is insufficient. In both the in-memory and disk-resident components, the interaction between learned models and LSM-specific access patterns introduces new challenges that limit performance and feasibility.\shorten{}

\subsection{Cold-Start Problem in Memtables}
\label{subsec:cold_start_problem}

\begin{figure}[t]
    \centering
    \includegraphics[width=1.04\linewidth]
    {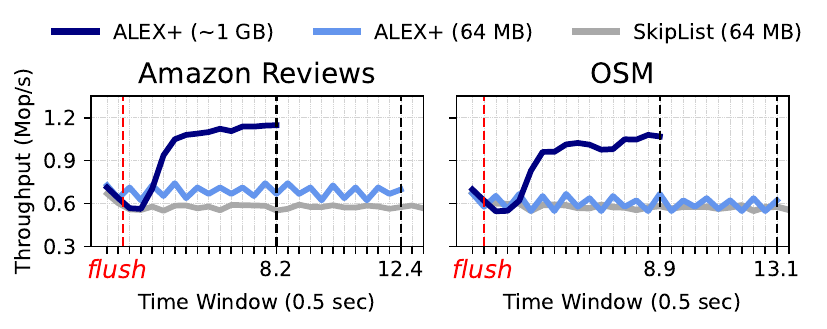} 
    \caption{Cold-start behavior of updatable learned indexes. Throughput per interval over 10M inserts for two data distributions: Amazon Reviews (near-linear) and OSM (non-linear). \alexol{} with a large memory budget (1 GB) remains unflushed, delivering high throughput; \alexol{} with a tight    budget (64 MB) is subject to frequent flushing and repeatedly incurs cold-start overheads resulting in degraded throughput. \shorten{}}
      \label{fig:cold_start}
\end{figure}


We first explore an out-of-the-box replacement of \rocksdb{}'s \skiplist{} Memtable with \alexol{}. Figure~\ref{fig:cold_start} compares the throughput of \alexol{} and \skiplist{} in a write-only workload on two data distributions. The experimental setup is the same as described in Section~\ref{sec:experiments}. We consider two memory configurations for \alexol{}. In the first, \alexol{} is given a 1 GB memory budget, to emulate conditions similar to those used in prior learned index studies \cite{gre, sosd, learned-index}. In the second, we constrain \alexol{} to 64 MB, a typical Memtable size in production systems~\cite{flodb, memtable_configuration}. Whenever the Memtable reaches its capacity, it is discarded and replaced with a new empty instance, emulating the standard LSM flush behavior. The \skiplist{} size is set at 64 MB

With a large memory budget, \alexol{} achieves up to $\sim$3$\times$ higher average write throughput than \skiplist{}, consistent with prior work \cite{learned_index_comparison_skiplist}. However, when deployed under realistic LSM constraints, end-to-end performance becomes comparable to or worse than \skiplist{}. This degradation stems from the fact that each new \alexol{} instance starts uninitialized and must undergo costly structural adaptations before reaching steady-state efficiency to absorb writes.


Unlike common learned-index benchmarks~\cite{sosd, sun2023learned}, Memtables are much smaller (typically 64-128 MB \cite{lsm,rocksdb}) and are short-lived. Once full, they are flushed to disk and replaced, leading to frequent rebuilds. These characteristics conflict with the assumptions of many updatable learned indexes, which incur significant initialization overhead and often rely on bulk-loading phases in which a significant fraction of the key space (e.g., up to 50\%) is inserted upfront for write-dominant workloads \cite{gre}. As a result, a learned index used as a Memtable repeatedly experiences cold starts, rebuilding its internal structure from scratch after each flush. 

Learned indexes such as ALEX~\cite{alex}, LIPP~\cite{lipp}, APEX~\cite{apex}, and XIndex~\cite{xindex} can outperform \skiplist{} in steady-state write-heavy workloads \cite{learned_index_comparison_skiplist}, but most assume such bulk-loaded initialization \cite{gre}. This is illustrated by the 1GB \alexol{}, which quickly converges to high throughput as it learns the key distribution, after an initial structure-adjustment phase. Under realistic memory limits, however, the Memtable is flushed roughly every $\sim$0.4 seconds (red dashed line), about three times more frequently than the convergence period of \alexol{}. This gap widens for more complex key distributions; for example, on OSM, flushing occurs up to five times more often than convergence, amplifying cold-start overheads. Additional examples of distribution difficulty are presented in Section~\ref{sec:experiments}.

To sum up, increasing Memtable size reduces cold-start frequency but raises flush costs and risks write stalls~\cite{balmau2019silk}. Consequently, addressing the problem requires improving index initialization rather than simply scaling memory. To this end, we introduce a reuse mechanism that preserves structural knowledge across Memtable instances, which we call skeletonization. This technique enables learned indexes to maintain their performance benefits despite frequent flush cycles and is described in detail in Section~\ref{subsec:skeletonization}.

\begin{figure*}[t]
    \centering
    \includegraphics[width=0.72\textwidth]{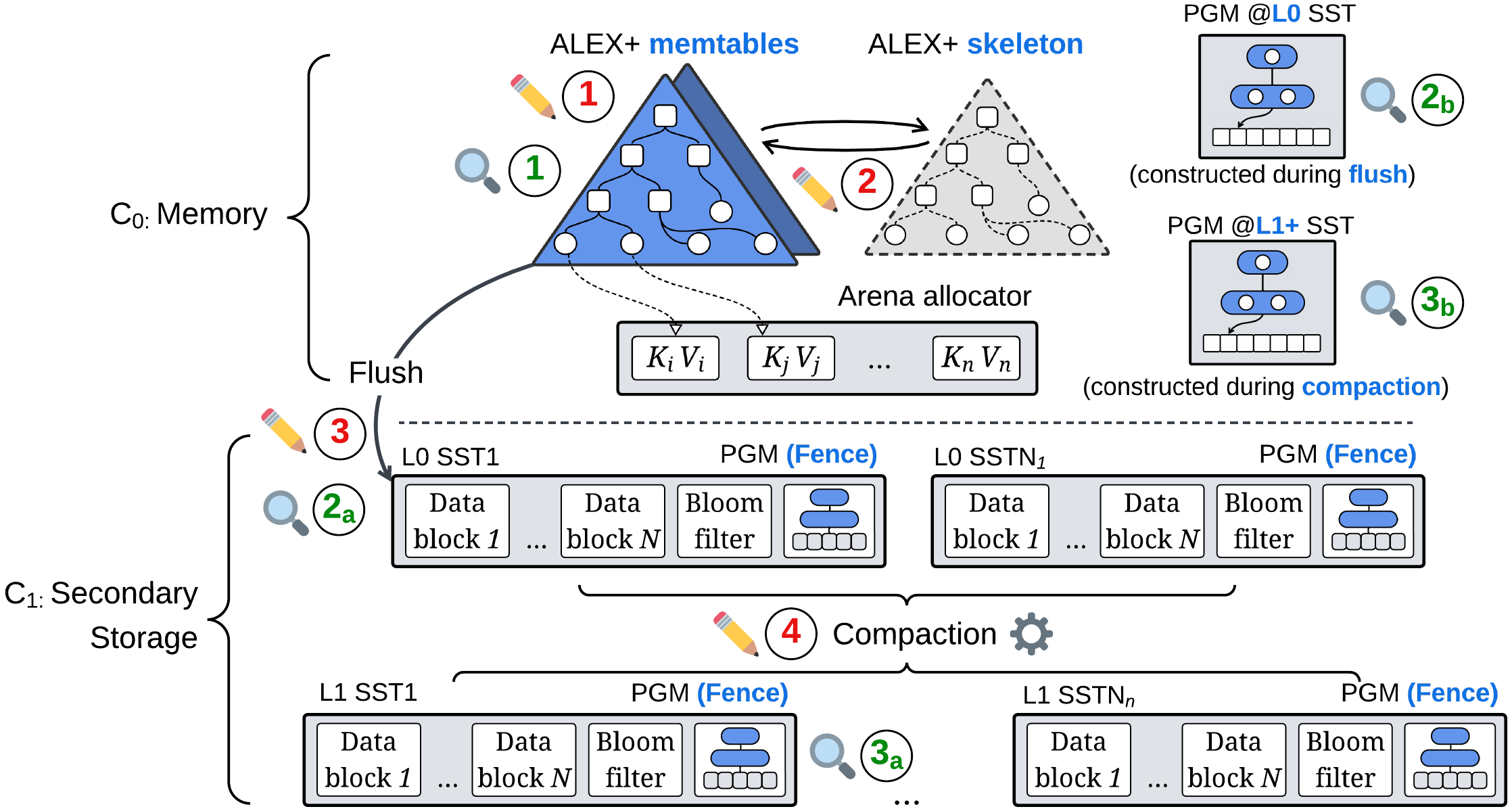} 
    \caption{\mountdb{}’s operational flow. The read and writes paths are virtually identical to those of RocksDB. \mountdb{} pushes the complexity of learned index maintenance into the LSM housekeeping tasks, flushing and compaction.}
    \label{fig:read_write}
\end{figure*}

\subsection{Challenges in Learning Disk-Resident Data}
\label{subsec:position_to_block}



\paragraph{Predicting record positions vs. blocks}
 Out-of-the-box in-memory learned indexes such as \pgm{} predict the position of a record in a logical contiguous array. In contrast, data on stable storage resides in files that are organized and transferred in units of data blocks or pages. Still, the same indexing principles can be applied to such  disk-resident data as long as every data block has the same number $n$ of records. In this case, we simply calculate the offset of the block in the file via $(block\_size * record\_ position) / n$. However, the true position of a record can differ as defined by the error bounds, and therefore could be in adjacent blocks.  If the record is not in the predicted block, neighboring blocks need to be retrieved. 
 
 This additional I/O cost can be avoided if all blocks within the error bound are retrieved in one I/O~\cite{pgm_relaxed_error_bound}. As these blocks are contiguous,  I/O will only be slightly more costly, in particular when error bounds are tight. In fact, \pgm{} can work with a very small error bound of $\varepsilon = 1$, which makes bulk I/O an attractive option. Another option is to use block-aligned variants of \pgm{} \cite{pgm_relaxed_error_bound} so that records are guaranteed to reside in the predicted block. 

\paragraph{Supporting variable record and block sizes } The challenge is that \rocksdb{} (and other popular LSM implementations such as LevelDB \cite{leveldb}, SpeedDB \cite{speedb}, and CockroachDB \cite{cockroachdb}) does not meet the assumptions required for these techniques: it stores variable-sized records, resulting in a variable number of entries per block, and its block sizes may themselves vary, especially under per-block compression. As a result, indexes such as \pgm{} cannot be used directly, even with the optimizations above.
What is required instead is an index that predicts the block of a record without relying on record sizes, block sizes, or intra-block offsets.

Several LSM-based designs move in this direction. Systems such as Bourbon \cite{bourbon}, LeaderKV \cite{leaderkv}, TridentKV \cite{tridentkv}, and DobLIX \cite{doblix} all train their models to predict the correct block. However, each comes with limitations that make them unsuitable for our goals. Bourbon depends on WiscKey~\cite{wisckey}, which separates fixed-length keys from values and therefore assumes predictable offsets in a key file. LeaderKV replaces SSTs entirely with a new storage format, while TridentKV introduces its own partitioning scheme. TridentKV and DobLIX both design a custom data block format. Moreover, they often rely on large blocks to reduce the likelihood of prediction errors, but such oversized blocks can lead to lengthy I/O operations and memory contention. In short, all of these systems modify the SST or block layout, whereas our objective is to keep \rocksdb{}’s SST format and block serialization unchanged. We seek only to replace the Fence Pointer index with a learned alternative. 

We therefore introduce PGM (Fence), a block-aligned variant of \pgm{} that preserves the advantages of the original model while guaranteeing block-level predictions. PGM (Fence) is highly compact because it trains on only one key per block—an idea inspired by Aulid \cite{aulid}, a hybrid learned index for traditional RDBMSs (see Section~\ref{sec:background}). PGM (Fence) ensures worst-case single-I/O lookup, supports variable-sized records and compressed blocks, and operates directly on unmodified \rocksdb{} SSTs and data block layouts.
Section~\S\ref{subsec:pgm_fence} details its design and integration into \mountdb{}.

\section{\mountdb{} Design \& Optimizations}
\label{sec:mountdb_design}

This section describes how \mountdb{} serves write and read operations, with minimal modifications to the classic LSM workflow.

\subsection{\mountdb{} Operation Flow}

While \mountdb{} is generally compatible with LSM key-value stores, we illustrate our techniques in an implementation extending RocksDB. The write and read paths in \mountdb{} largely follow the conventional LSM workflow, reusing most of the RocksDB logic. Figure~\ref{fig:read_write} illustrates these operations across the LSM hierarchy. 

\vspace{1mm}
\noindent\textbf{Write path.} 
In \rocksdb{}, writes insert a new key–value pair into the Memtable {\Large \ding{172}}. The underlying memory for key–value storage is managed by the arena allocator, which allocates contiguous chunks, and keys are appended sequentially in insertion order. Additional meta-information builds a \skiplist{} that connects records according to their keys. This \skiplist{} is traversed to quickly find records with specific key values.  In \mountdb{}, we replace this meta-information with an \alexol{} index where the leaf nodes, instead of storing full key-value pairs, maintain pointers to the actual records.

\vspace{1mm}
\noindent\textbf{Read path.} The read path in \mountdb{} is identical to that of RocksDB, with the exception that the RocksDB SST Fence Pointer Table index is replaced with a \pgmfence{} index. Like in RocksDB, lookup requests traverse the LSM hierarchy top-down. The search begins with probing the active \alexol{} Memtable, followed by the immutable Memtable {\Large \ding{172}}. If the key is not found in memory, the search proceeds to $L_0$. Bloom filters quickly eliminate irrelevant files {\Large \ding{173}}$_a$. If a Bloom filter indicates presence, the corresponding SST index is loaded into the OS page cache and used to determine the target data block, which is then also loaded into memory, and a binary search is performed to locate the key {\Large \ding{173}}$_b$. For subsequent requests, the index remains cached in the OS page cache. For deeper levels (L$_1$, L$_2$, …), the same process is repeated: {\Large \ding{174}}$_a$ Bloom filters are checked, and {\Large \ding{174}}$_b$ the SST index guides access to the target block.

\vspace{1mm}
\noindent\textbf{Background operations: flushing and compaction.} A key feature of \mountdb{}'s design is pushing the complexity of learned index creation and maintenance off the critical path, in the LSM housekeeping operations. During flushing in RocksDB, when the active Memtable reaches its memory budget, it is marked immutable, and a new Memtable is created {\Large \ding{173}}. In \mountdb, we additionally apply skeletonization to initialize the new Memtable using the structure of the previous immutable instance, mitigating cold-start inefficiencies (see Section~\ref{subsec:skeletonization}). Note that we perform skeletonization only periodically, reusing the same skeleton for several new Memtables to reduce overhead.
Next, the immutable Memtable is flushed to disk, producing the SST’s data blocks. Once the data portion of the SST is materialized, instead of creating the conventional \rocksdb{} Fence Pointer Table, \mountdb{} constructs the \pgmfence{} index over the SST’s fence keys (see Section~\ref{subsec:pgm_fence}). Because the keys are already sorted, index construction is efficient. The \pgmfence{} index and the existing Bloom filter from \rocksdb{} are then appended as metadata, finalizing the SST, which is persisted to disk {\Large \ding{174}}.

Compaction follows the same principle as in RocksDB. When SSTs at $L_0$ are merged, their key ranges are combined to produce the data blocks of a new SST. Once the merged data is materialized, \mountdb{} constructs a new \pgmfence{} index over the resulting fence keys. The index is then appended to the SST as metadata, and the finalized file is persisted to disk {\Large \ding{175}}. 

In this work, we focus on numeric keys. Learned indexing techniques for strings have been proposed in prior work \cite{lits, slipp, sindex, rss, tridentkv, doblix, leaderkv}. However, they are often specifically designed for strings~\cite{lits, slipp, sindex, rss, doblix} and thus, are orthogonal to learned indexes for numeric keys, or require changing the storage format of the records~\cite{tridentkv}. Therefore, integrating these approaches with \mountdb{}’s architecture is a promising direction that we leave to future work.



\subsection{Solving the Cold-Start Problem with Skeletonization}
\label{subsec:skeletonization}

\begin{algorithm}[t]
\caption{Skeletonization of  \alexol{} Memtables}
\label{alg:skeletonization}
\begin{flushleft}

\textbf{Input:} $\phi$: skeletonization frequency, $\mathit{Memtable}_{\mathit{size}}$: memory budget 
\textbf{Initialize:} $\mathcal{S}$: global skeleton ($\mathcal{S} \leftarrow \texttt{NULL}$), $c$: flush counter ($c \leftarrow 0$)
\end{flushleft}
\begin{algorithmic}[1]
    \State $M_{\text{active}} \leftarrow \text{ALEX+}()$ \small{\Comment{initialize first Memtable from scratch}}
    \For{each write operation} \do \\
        \State $M_{\text{active}}.\textsc{Insert}(k, v)$
        \If{$M_{\text{active}}.\textsc{Size}() \geq \mathit{Memtable}_{\mathit{size}}$}
            \State $M_{\text{immutable}} \leftarrow M_{\text{active}}$ \small{\Comment{read-only}}
            \State $M_{\text{active}} \leftarrow \textsc{CreateNewMemtable}(\mathcal{S})$
            \State $c \leftarrow c + 1$
            \If{$c \bmod \phi = 0$}  \small{\Comment{create skeleton in background}}
            \State $\textsc{ScheduleBackgroundTask}(\textsc{CreateSkeleton},$
            \hspace*{1.18cm} $M_{\text{immutable}})$
            \EndIf
        \EndIf
    \EndFor

    \Statex
    \Function{CreateNewMemtable}{$\mathcal{S}$}
        \If{$\mathcal{S} = \texttt{NULL}$} \small{\Comment{no skeleton available (initial case)}}
            \State \Return $\text{ALEX+}()$
        \Else
            \State \Return $\text{ALEX+}().\textsc{CopySkeleton}(\mathcal{S})$ 
        \EndIf
    \EndFunction

    \Statex
    \Procedure{CreateSkeleton}{$M$}
        \State $K_{\text{sorted}} \leftarrow M.\textsc{GetSortedKeys}()$ \small{\Comment{extract sorted keys}}
        \State $M_{\text{temp}} \leftarrow \text{ALEX+}().\textsc{BulkLoad}(K_{\text{sorted}})$  \label{line:bulkload} \small{\Comment{build optimal structure}}
        \For{each dataNode $d$ in $M_{\text{temp}}.\textsc{DataNodes}()$} \do \\
            \For{$i \gets 0$ \textbf{to} $d.\textit{dataCapacity}$} \do \\
                \State $d.\textit{payloadSlots}[i] \leftarrow \texttt{nullptr}$
                \label{line:cleaning_one}
                \small{\Comment{clear payloads}}
            \EndFor
            \State $d.numInserts \leftarrow 0$
            \small{\Comment{reset stats}}
            \State $d.numLookups \leftarrow 0$
            \State $d.numShifts \leftarrow 0$ 
            \State $d.numResizes \leftarrow 0$
            \State $d.\textsc{ResetLock}()$
            \label{line:cleaning_two}
        \EndFor
        \State $M_{\text{temp}}.\textsc{RemovePayloads}()$ 
        \State $\mathcal{S} \leftarrow  M_{\text{temp}}.\textsc{CopyModelNodes}()$ 
        \label{line:selective_copy_one}
        \small{\Comment{includes models and links}}
        \State $\mathcal{S} \leftarrow  M_{\text{temp}}.\textsc{CopyDataNodes}()$
        \label{line:selective_copy_two}
        \small{\Comment{includes gapped arrays \& keys}}
    \EndProcedure
\end{algorithmic}
\end{algorithm}

To make \alexol{} effective as a short-lived Memtable, \mountdb{} introduces a novel \textit{skeletonization} mechanism that reuses the structure of previous Memtables. As mentioned in Section~\ref{sec:design_tensions}, one of the assumptions when using learned indexes is that their initialization cost would be amortized over a long usage period. On a high level, skeletonization effectively freezes the internal structure of the learned index, assuming that the key distribution patterns evolve slower than the rate of flushing of the Memtable.

Algorithm \ref{alg:skeletonization} describes how \mountdb{} applies skeletonization to \alexol{} Memtables. Initially, the first two Memtables are created from scratch and are empty. Once the first Memtable reaches its budget and is marked as immutable, it is flushed to disk while a new active Memtable begins absorbing inserts. In the background, \mountdb{} invokes the standard bulk-load operation of \alexol{} on the immutable Memtable’s sorted keys (line~\ref{line:bulkload}) to construct an optimized \alexol{} instance. \shorten{}

From this bulk-loaded instance, \mountdb{} extracts a skeleton that retains only the structural components defining the index layout. This is achieved through a selective deep copy of the bulk-loaded \alexol{} instance (see lines \ref{line:selective_copy_one}-\ref{line:selective_copy_two}). The skeleton preserves the following: 1) a complete hierarchy of model and data nodes, 2) all parent–child relationships, 3) inter-node links that connect adjacent data nodes (used for leaf-level iteration), and 4) key-placement metadata such as the keys and gaps in the gapped arrays. Importantly, for gapped arrays, \mountdb{} maintains the exact spatial pattern of gaps by preserving the bitmap representation that marks occupied versus free slots within each array. Consequently, the model and data nodes preserve the learned key distribution from the previous workload, while the preserved gaps capture the fine-grained insertion topology that emerged from prior Memtable creation. In contrast, all data-dependent elements, such as payloads, statistical counters for number of inserts, lookups, key shifts and node resizes, along-with lock states, are reset (described in lines \ref{line:cleaning_one}-\ref{line:cleaning_two}), delivering a clean skeleton for reuse.

Effectively, the skeleton preserves the expensive structural foundation of the previous Memtable. When a new Memtable is created, it is initialized from this skeleton, allowing new insertions to immediately leverage the optimized index layout. During insertion, if the target key position in the gapped array already contains an existing key, \mountdb{} inspects its corresponding payload pointer. 
A \texttt{nullptr} indicates that the slot is free, the existing key can be overwritten with the new key, and the payload pointer is updated to reference the new value. This reuse of pre-allocated gaps and key slots enables efficient inserts without triggering structural modifications or reorganization. As a result, \mountdb{} achieves efficient writes while preserving the learned model hierarchy and spatial layout of previous Memtables.

\mountdb{} does not create a new skeleton after each Memtable flush but instead reuses the same skeleton for consecutive Memtables. To adapt to shifting key distributions, \mountdb{} introduces a parameter $\phi$ that determines how frequently (in terms of number of flushes) the skeletons are refreshed. Based on empirical evaluation, we set $\phi = 10$. All skeleton creation and maintenance are performed during the flush operation, off the critical path, ensuring that foreground write throughput remains unaffected. The overhead introduced by skeletonization is minimal, discussed in Experiment~\ref{sec:compaction-eval}, accounting for only  $\sim$2.1\% of the time of an average flush operation.

\subsection{\pgmfence{}: Maintaining a Single I/O per Lookup with Variable Size Records}
\label{subsec:pgm_fence}

As explained in Section~\ref{sec:design_tensions}, our objective is to integrate a learned index at the disk-level with minimal modification to the existing LSM storage structure. In particular, \mountdb{} only replaces the fence pointer table in each SST while keeping the data block structure unchanged, reusing \rocksdb{}'s efficient binary search on variable length records inside the block.


Given a query key, the learned index needs to find the offset of the block in the SST that contains the corresponding record.  A crucial concept in our approach is that we do not train the index over all keys in the SST but only over the key of the first record of each block~\cite{aulid}. Whether a query requests this first record or any other record in a block $b$, the index will estimate $b$'s offset or the offset of one of $b$'s adjacent blocks. This reduction in training records reduces not only learning time but also index size. 


\begin{figure}[t]
    \centering
    \includegraphics[width=1.01\linewidth]
    {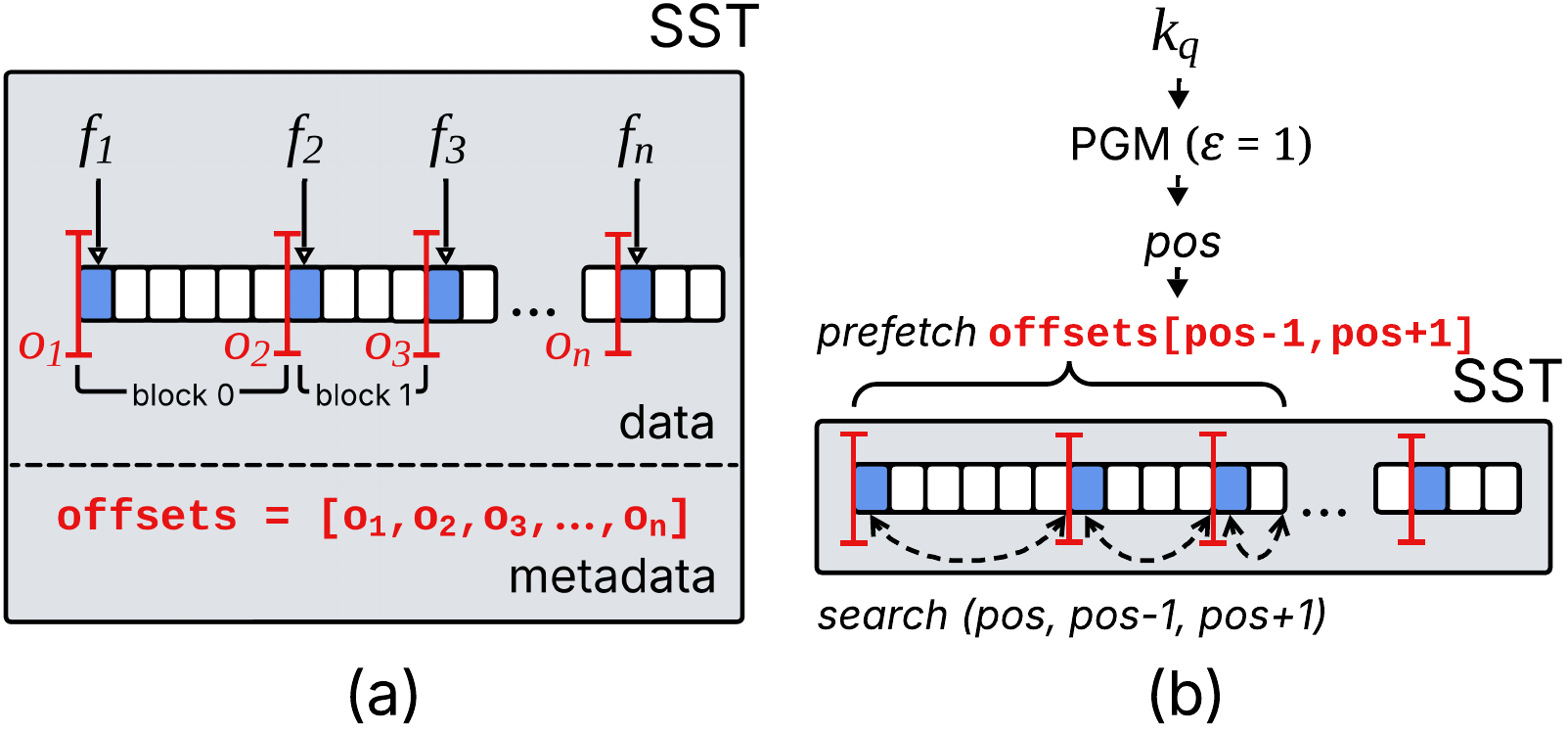} 
    \caption{(a) Fence-Key Modeling \& (b) Lookup in \pgmfence{}\shorten{}} 
      \label{fig:pgm_fence}
\end{figure}

\begin{algorithm}[t]
\caption{PGM (Fence) Index: Construction and Lookup}
\label{alg:pgm_fence}
\begin{flushleft}

\textbf{Input:} $\mathcal{S}$: SST with sorted sequence of blocks $\{B_1, B_2,\dots, B_N\}$,\\ $k_q$: lookup key  \\
\textbf{Initialize:} $\mathcal{P}_{\textit{fence}}$: PGM (Fence) Index ($\mathcal{P}_{\textit{fence}} \leftarrow \texttt{NULL}$), $o$: offsets array ($o \leftarrow \emptyset$)

\end{flushleft}
\begin{algorithmic}[1] \small{}
    \Function{Train}{$\mathcal{S}$}
        \State $f$: fence-keys array ($f \leftarrow \emptyset$)
        \For{$\textit{block} \in \mathcal{S}$} 
        \label{line:pass_over_SST}
        \do \\
            \State $f[i] \leftarrow \textit{block} .\textit{firstKey}$ 
            \State $o[i] \leftarrow \textit{block} .\textit{blockOffset}$ 
        \EndFor
            \State $\mathcal{P}_{\textit{fence}} \leftarrow \textsc{TrainPGM}(f, \varepsilon = 1)$
            \label{line:train_pgm_fence}
            \State $f \leftarrow \emptyset$ \small{\Comment{deallocate fence-key array}}
        \State \Return $\mathcal{P}_{\textit{fence}}$ 
    \EndFunction

    \Statex
    \Function{Lookup}{$k_q$}
    \State $p \leftarrow \mathcal{P}_{\textit{fence}}.\textsc{predict}(k_q)$ \small{\Comment{position in $f$ (search window $= 2\varepsilon+1$)}}
    \State $idx_{\textit{start}} \leftarrow \max(0, p - 1)$
    \State $idx_{\textit{end}} \leftarrow \min(o.\textit{length}-1, p + 1)$
    \State $\textit{byte}_{\textit{start}} \leftarrow o[idx_{\textit{start}}]$
    \State $\textit{byte}_{\textit{end}} \leftarrow  (idx_{\textit{end}}+1<o.\textit{length}) \,?\:o[idx_{\textit{end}}+1] : \texttt{EOF}$ 
    \State $\mathcal{B}_{\textit{mem}} \leftarrow \textsc{Prefetch}(\mathcal{K}, \textit{byte}_{\textit{start}}, \textit{byte}_{\textit{end}})$ 
    \small{\Comment{single I/O}}
    \label{line:prefetch}
    \For{$\textit{block} \in \mathcal{B}_{\textit{mem}}[p, p-1, p+1]$} \do \\ \small{\Comment{prioritize  block order}}
    \label{line:search_one}
        \State $(\texttt{Found},\textit{record}) \leftarrow \textsc{binarysearch}(\textit{block}, k_q)$ 
        \If{$\texttt{Found}$}  
            \State \Return $\textit{record}$
        \EndIf    
        \label{line:search_two}
    \EndFor
    \State \Return \texttt{NotFound}    
    \EndFunction

\end{algorithmic}
\end{algorithm}



Our index \pgmfence{} is a block-aware adaptation of the \pgm{} index. Algorithm~\ref{alg:pgm_fence} illustrates the construction and lookup procedure. After an SST is materialized, \mountdb{} performs a single pass over the block metadata to extract from each data block (line \ref{line:pass_over_SST}): (i) the key of the first record in the block, which we refer to as  \textit{fence key}, and (ii) the  byte offset of the block. The fence keys are stored in a temporary array $f$, which is discarded at the end of the learning process. The offsets are stored in an offsets array $o$. Both are illustrated in Figure~\ref{fig:pgm_fence}(a). Note that both arrays are of same length, with $f[i]$ storing the first key of block $i$ and $o[i]$ the offset of block $i$.

We then train a \pgm{} solely on the ordered fence-key sequence $f$, i.e., \pgm{} will learn the position of each fence key in $f$. The \pgm{} model is trained with the tightest error bound $\varepsilon=1$ (line \ref{line:train_pgm_fence}), enabling the index to exploit the model’s error guarantees and limit the lookup window to at most 3 positions (blocks).  After training, the temporary fence-key array is discarded, and only the learned model, together with the offsets array, is stored alongside the SST.



At lookup time, the \pgm{} model predicts the position $p$ of the query key $k_q$ within the fence-key sequence $f$ (line 12). Since each fence key uniquely represents a block boundary, the predicted position $p$ directly maps to the same position $p$ in the corresponding offsets array $o$, which will give us the Byte offset of the estimated block for $k_q$. Since the error bound $\varepsilon=1$, the record will be in the estimated block or its left resp.\ right neighbor block (lines 13-14).  Therefore, only a small window of candidate blocks needs to be examined, which is three in this case, as shown in Figure~\ref{fig:pgm_fence}(b). The corresponding byte ranges are then prefetched in a single request using \rocksdb{}’s \texttt{FilePrefetchBuffer} (line \ref{line:prefetch}), preserving \rocksdb{}’s default block-oriented access pattern without modifying the storage layout.
Typically, LSM datastores have the block size configured to a multiple of 4KB, to take advantage of the access granularity of SSDs. To avoid I/O amplification caused by retrieving 3 blocks instead of a single block, \mountdb{} configures the default maximum block size of the LSM datastore to a third of its default. 
Within the prefetched region, the final key lookup is performed using \rocksdb{}’s existing binary search inside each block. 

The predicted (middle) block is binary-searched first, followed by the left and right neighbors if necessary (lines~\ref{line:search_one}–\ref{line:search_two} in Algorithm~\ref{alg:pgm_fence}). Our empirical observations show that approximately 78\% of positive lookups are resolved in the predicted block, with 14\% and 8\% in the left and right neighbors, respectively, making this search ordering effective in minimizing lookup latency.


A final advantage of indexing on the block start as opposed to over the entire key range is the significant reduction in index size. The learned \pgm{} model together with the offsets array take less than 172KB per SST even under the most challenging key distributions we evaluated. While this is not significant inside each SST (as their default size is in the order of tens of MBs, e.g., 64MB), we show that the small \pgm{} size provides an advantage in read-intensive workloads. As the index is small, more records can be cached within the same memory budget, leading to higher read throughput.

\section{Evaluation}
\label{sec:experiments}

\begin{figure*}[t]
    \centering
    \includegraphics[width=\textwidth]{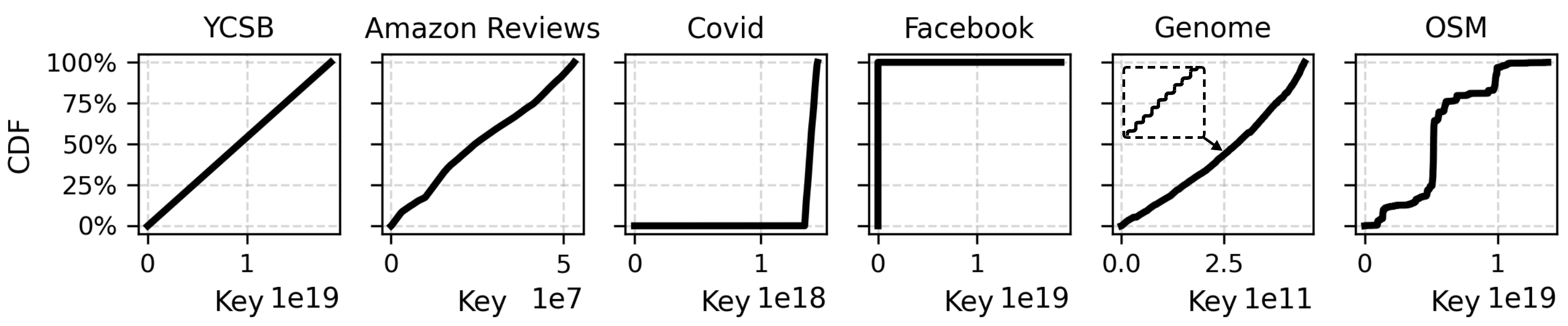} 
    \caption{Cumulative distribution function (CDFs) of real-world datasets, ordered by increasing indexing difficulty (left to right)} 
    \label{fig:cdfs}
\end{figure*}

\begin{figure*}[t]
    \centering
    \includegraphics[width=\textwidth]{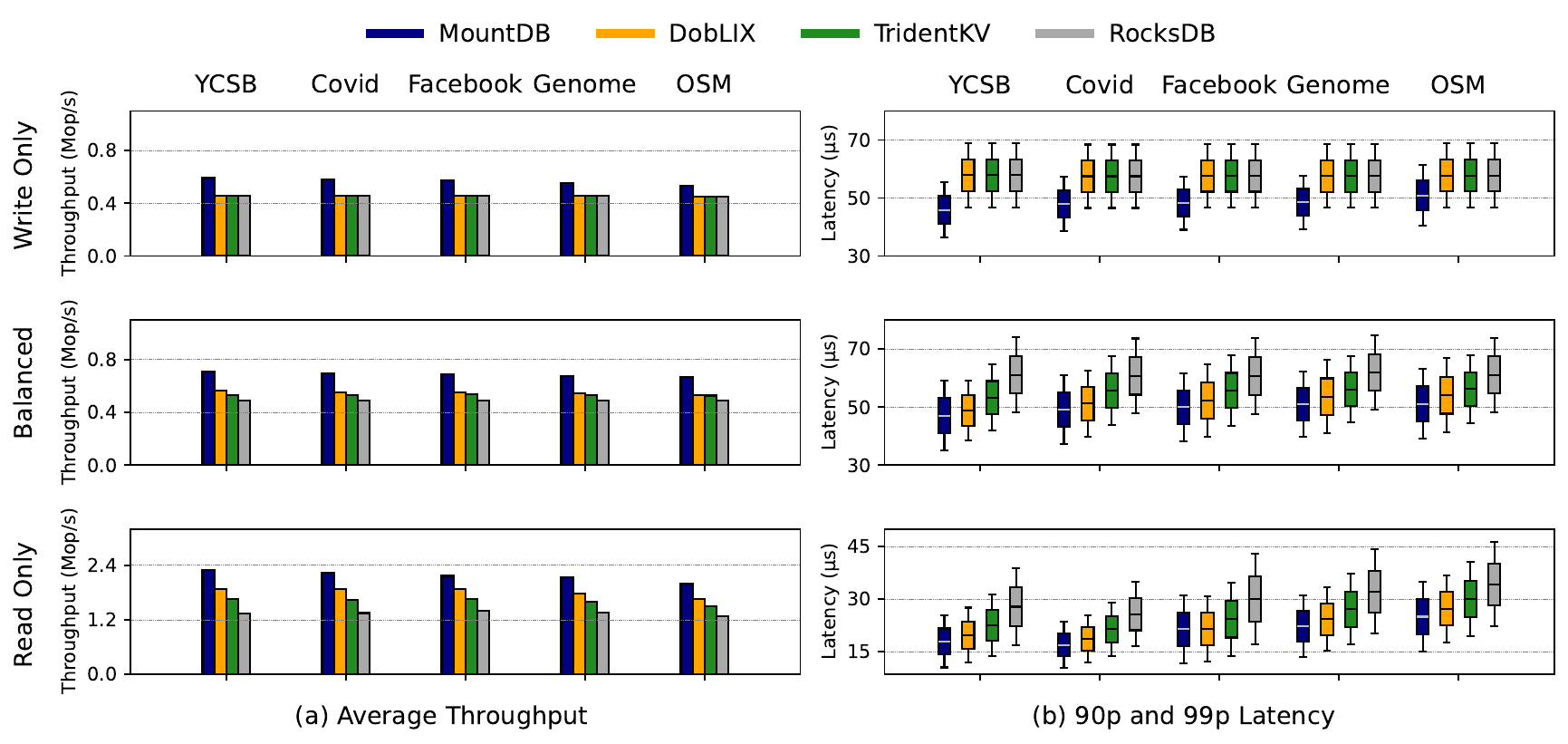} 
\caption{\textbf{Large-Scale Workloads.} (a) Average throughput and (b) 90th and 99th percentile latencies across five datasets, each  $\sim$55 GB, with 16 application threads. Results are shown for write-only (top), balanced (50$\%$ reads / 50$\%$ writes, middle), and read-only (bottom) workloads under a uniform access pattern.\shorten{}}
    \label{fig:exp_large_scale_concurrent}
\end{figure*}

We structure our evaluation around four main questions: 
\begin{itemize}[leftmargin=*, nosep]
\item What is \mountdb{}'s end-to-end performance with different read-to-write ratios and key distributions? (Section~\ref{sec:eval-end-to-end})

\item Can \mountdb{} adapt to changing key distributions over time, especially on the write path? (Section~\ref{sec:skeleton-eval})

\item Do \mountdb{}'s indexes provide a latency advantage on the read path, especially for disk-resident data? (Section~\ref{sec:read-performance-eval})

\item How much overhead do learned indexes introduce in LSMs, particularly to flush and compaction operations? (Section~\ref{sec:compaction-eval})

\end{itemize}

\subsection{Experimental Setup}

\noindent \\ \textbf{Hardware.} All experiments were conducted on a machine running Ubuntu 22.04.5 LTS, equipped with two Intel Xeon E5-2690 v4 CPUs, each with 14 cores running at 2.60 GHz, 126 GB of RAM, and a 500 GB NVMe SSD for storage. 
\vspace{1mm}
\noindent \\ \textbf{Baselines.}
We compare \mountdb{} against three LSM-based baselines: two state-of-the-art learned indexing systems\footnote{The publicly available code for \doblix{} \cite{doblix-code} and \tridentkv{} \cite{tridentkv-code} uses older versions of \rocksdb{} (5.1.2 and 5.4.10, respectively from 2017). For fairness, we ported both systems to \rocksdb{} v8.10.0. Additionally, the \doblix{} repository does not include the RL-based tuner described in the paper
.}, \doblix{} and \tridentkv{}, and a production-grade LSM, \rocksdb{} (v8.10.0). All baselines are configured identically to \mountdb{} in terms of Memtable size, cache allocation, and compaction resources. All experiments are run through the standard benchmarking tool $\texttt{db\_bench}$. \bourbon{}~\cite{bourbon}, the first system to add learned indexes into an LSM, is omitted as prior work~\cite{doblix} shows that \doblix{} and \tridentkv{} outperform it. Additionally, LeaderKV \cite{leaderkv} is not open source.\shorten{}


\vspace{1mm}
\noindent \\ \textbf{Datasets.} We use six datasets with diverse real-world key distributions. Figure~\ref{fig:cdfs} shows the datasets ordered by increasing indexing difficulty~\cite{gre}. Amazon Reviews contains 21.9M unique keys, while each of the other datasets contains 200M unique keys. Facebook and OSM appeared in the SOSD benchmark~\cite{sosd}, YCSB in \alexol{}'s evaluation~\cite{alex,cooper2010benchmarking}, Amazon Reviews in \bourbon{}'s evaluation~\cite{bourbon}, Covid and Genome appear in the GRE benchmark~\cite{gre}. For  read:write ratios we use the YCSB~\cite{cooper2010benchmarking} benchmark suite with the 6 default access patterns. \shorten{}


\vspace{1mm}
\noindent \\ \textbf{System configuration.}
Unless otherwise specified, we use  default \rocksdb{} configuration with two 64 MB Memtables (one active and one immutable) and a 32~MB block cache, reserved exclusively for data blocks. For large-scale experiments we use a 2~GB block cache. Compactions are executed with 8 worker threads. Compression is disabled. We run all workloads with 16 client threads. For all experiments except those that analyze the impact of value size and test performance under variable KV pairs, records consist of 8-byte keys and 100-byte values. Each experiment is repeated three times, and we report the average across runs. \shorten{}

\subsection{End-to-end Results}
\label{sec:eval-end-to-end}

\begin{figure*}[t]
    \centering
    \includegraphics[width=1.01\textwidth]{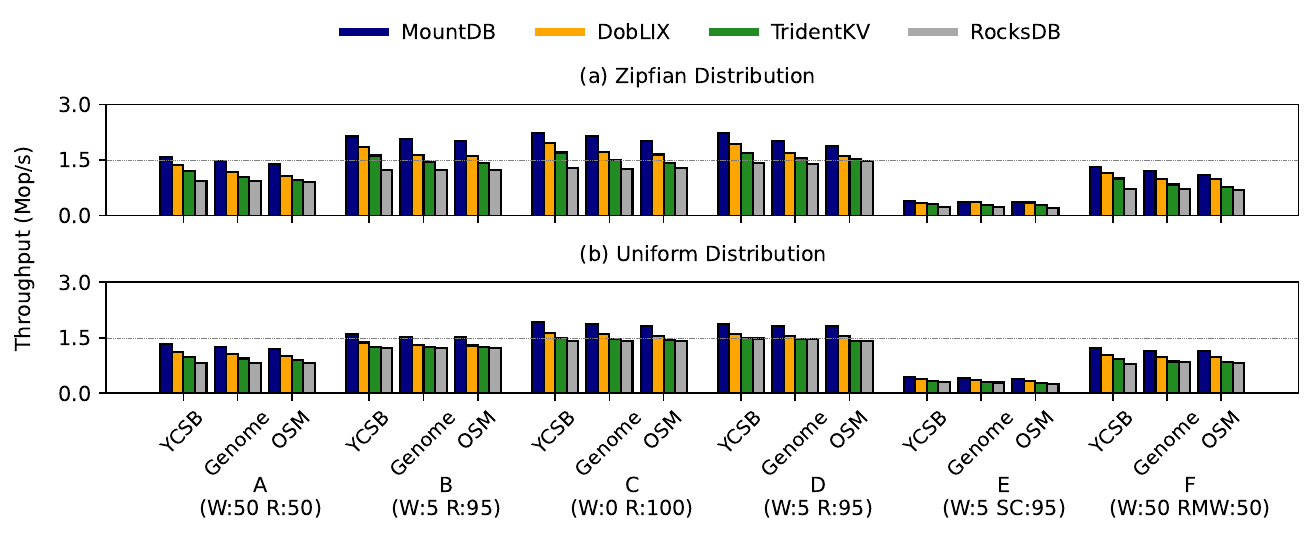}
    \vspace{-6mm}
    \caption{Throughput comparison across the six YCSB workloads (A–F) and three datasets (YCSB, Genome, OSM) under (a) Zipfian and (b) Uniform access distribution.}
    \label{fig:exp_ycsb}
\end{figure*}

\vspace{1mm}
\noindent\textbf{Varying key distribution.} 
Figure~\ref{fig:exp_large_scale_concurrent} reports the average throughput (Figure~\ref{fig:exp_large_scale_concurrent}(a)) and tail latencies (Figure~\ref{fig:exp_large_scale_concurrent}(b)) across write-only, balanced (1:1 read–write ratio), and read-only workloads on multiple datasets with different key access patterns. Each dataset is approximately 55 GB in size, and each run executes 500M operations. \shorten{} 

For write-only workloads (top row), \mountdb{} achieves up to 1.3$\times$ higher throughput against \skiplist{}, which is used by all other baselines. 
Even on the most challenging dataset (OSM), the throughput improves by around 1.2$\times$. Since write-heavy workloads primarily stress the in-memory components of the LSM, these improvements stem from the optimized Memtable index and the use of skeletonization to mitigate cold-start overheads in updatable learned structures. In contrast, the other baselines retain the default \skiplist{}-based Memtable implementation and therefore exhibit performance comparable to \rocksdb{}, as they do not modify or optimize the in-memory indexing layer.

The balanced workload (middle row) shows consistent throughput improvements across datasets, reaching up to 1.3$\times$ over \doblix{}, 1.4$\times$ over \tridentkv{}, and 1.44$\times$ over \rocksdb{}. These gains arise from the combined effect of the in-memory \alexol{} index and the disk-resident \pgmfence{} index. Because lookups first probe the Memtables, \alexol{} improves both update efficiency and the latency of recent reads, while \pgmfence{} accelerates access to keys that are not resident in memory by narrowing the block search space. Read-only workloads exhibit the largest relative improvements, with throughput gains of up to 1.29$\times$ compared to \doblix{}, 1.58$\times$ compared to \tridentkv{}, and 1.77$\times$ compared to \rocksdb{}. 

Latency reductions at the 90th and 99th percentiles are consistent with the throughput improvements. For write-only workloads, \mountdb{} reduces tail latencies by 1.3$\times$ on YCSB and achieves smaller but consistent reductions on harder datasets (1.1$\times$). Balanced and read-only workloads exhibit a similar ordering across systems, with \mountdb{} and \doblix{} consistently showing lower tail latencies. This behavior stems from the use of learned indexing, which reduces lookup variability by narrowing the search space and, in most cases, limiting the lookup path to a single disk I/O. In contrast, \tridentkv{} exhibits a wider tail distribution, due to variability in its block sizes and longer last-mile searches. 
These trends remain consistent even on more challenging datasets such as Genome and OSM, where the performance gap between \mountdb{} and \doblix{} becomes more pronounced, with \mountdb{} exhibiting consistently smaller tail latencies.


\noindent\textbf{Varying read-write patterns.}
The goal of this experiment is to evaluate performance across a wide range of workload mixes and access patterns using the YCSB benchmark suite. 
We consider three datasets with distinct key distributions (YCSB--easy, Genome--medium difficulty, and OSM--hard) and the six YCSB workloads: A (update-heavy), B (read-mostly), C (read-only), D (read-latest), E (scan-heavy), and F (read-modify-write). Following the standard YCSB setup, each workload is evaluated under both Zipfian and uniform access distributions.

Figure~\ref{fig:exp_ycsb} reports the average throughput across these settings. Across all workloads and datasets, \mountdb{} consistently achieves the highest throughput. In update-heavy workloads (A and F), throughput improvements reach up to 1.2–1.3$\times$ over \doblix{}, 1.3–\\1.4$\times$ over \tridentkv{}, and around 1.4$\times$ over \rocksdb{}, reflecting the efficiency of the optimized write path of \mountdb{}. 

For read-dominant workloads (B, C, and D), \mountdb{} maintains clear performance advantages, with throughput gains typically in the range of 1.17–1.3$\times$ over \doblix{} and 1.2–1.4$\times$  over \tridentkv{}. 
These improvements arise from the combined effect of skeletonized \alexol{} and \pgmfence{}, which together reduce lookup overhead across both hot and cold data. In the scan-heavy workload (E), \mountdb{} continues to outperform all baselines, with gains of up to 1.25$\times$ , demonstrating that narrowing the search space at the block level also benefits workloads that access contiguous key ranges.

Access distribution influences performance trends but does not change the relative ordering. Under Zipfian access patterns (Figure~\ref{fig:exp_ycsb}, top-row), the performance gap widens slightly due to improved cache locality and faster adaptation of learned indexes. 
Across datasets, smoother key distributions (e.g., YCSB) yield larger improvements, while more irregular distributions such as OSM narrow the margins but still show consistent gains, highlighting \mountdb{}'s robustness across diverse data distributions.


\vspace{1mm}
\noindent\textbf{Performance under Varying and Mixed Value Sizes.} 

\begin{figure}[t!]
    \centering
    
    \begin{subfigure}{\linewidth}
        \centering
        \includegraphics[width=1.04\linewidth]{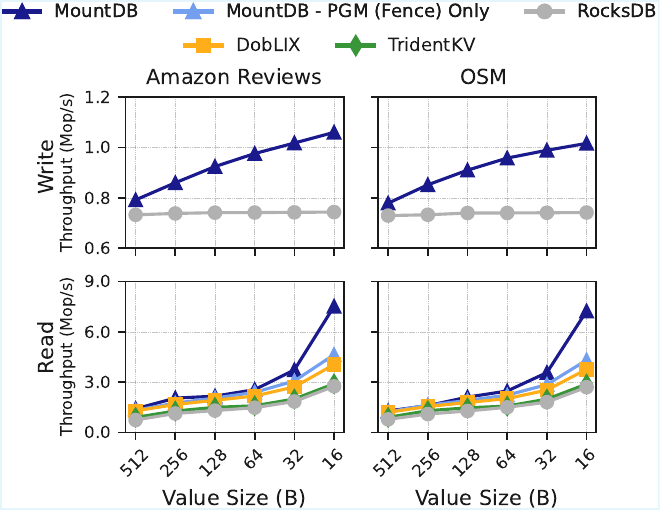}
        \captionsetup{labelfont=normalfont,textfont=normalfont}
        \caption{Scaling Value Size}
        \label{fig:exp_scaling_value_size}
    \end{subfigure}

    \begin{subfigure}{\linewidth}
        \centering
        \includegraphics[width=0.92\linewidth]{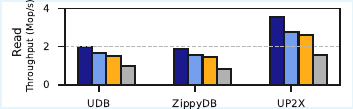}
        \captionsetup{labelfont=normalfont,textfont=normalfont}
        \caption{Mixed Value Size}
        \label{fig:example-b}
    \end{subfigure}
    
    \caption{\textbf{Impact of variable and mixed value size.} (a) Write (top) and read (bottom) throughput as value size decreases. (b) Performance under mixed value size workloads (UDB, ZippyDB, UP2X).}
    \label{fig:variable_kv}
\end{figure}

LSM systems are notorious for not being able to handle small value sizes well, with special LSM systems being proposed to mitigate this problem (such as PebblesDB~\cite{pebblesdb}). We show that using learned indexes can offer an elegant solution to this problem.
We first study the impact of value size on system performance, shown in Figure \ref{fig:variable_kv}(a). 
The key size is fixed at 8 bytes, while value size is reduced from 512 to 16 bytes. Memtable capacity remains fixed at 64~MB, meaning smaller values allow each Memtable to hold proportionally more records.

For writes, the \skiplist{} implementation used by the baselines exhibits nearly constant throughput across all value sizes on both datasets. This stability follows from its $\mathcal{O}(\log n)$ insertion complexity, which is largely insensitive to the number of stored key–value pairs~\cite{skiplist}. In contrast, \mountdb{} shows steadily increasing throughput as value size decreases. With smaller values, \alexol{} stores more records within the same Memtable capacity, providing additional training data for model refinement. This improves the model-to-data ratio, allowing each learned model to predict positions for a larger set of keys and thereby reducing lookup and update overheads. Consequently, write throughput improves as value size decreases showing up to 1.5$\times$ throughput improvements. 

For reads, throughput increases for all systems as value size decreases due to reduced I/O and improved cache residency. \mountdb{} benefits additionally from the combined effect of \alexol{} and the \pgmfence{} index. 
As a result, \mountdb{} achieves up to $2.8\times$ higher read throughput compared to \rocksdb{}, and up to 2.1$\times$ compared to \doblix{} for the smallest value sizes.

We also evaluate performance under mixed value size workloads derived from real-world \rocksdb{} deployments: UDB (avg value = 126.7 B, $\sigma$ = 22.1 B), ZippyDB (avg value = 42.9 B, $\sigma$ = 26.1 B), and UP2X (avg value = 46.8 B, $\sigma$ = 11.6 B) ~\cite{mixed_kv_size_workload}. 

Across all workloads, \mountdb{} consistently achieves the highest read throughput (see Figure \ref{fig:variable_kv}(b)). 
It outperforms \doblix{} by 1.2–1.4$\times$ across workloads and exceeds \rocksdb{} by 1.3–1.6$\times$, with the largest gains observed in UP2X. The \pgmfence{}-only configuration also improves throughput relative to the baselines, but the end-to-end \mountdb{} design consistently delivers larger improvements, reflecting the advantages of full-stack learned indexing. \shorten{}


\subsection{Adapting to Workload Shifts}
\label{sec:skeleton-eval}

\begin{figure*}[!t]
    \centering
    \includegraphics[width=1.01\textwidth]{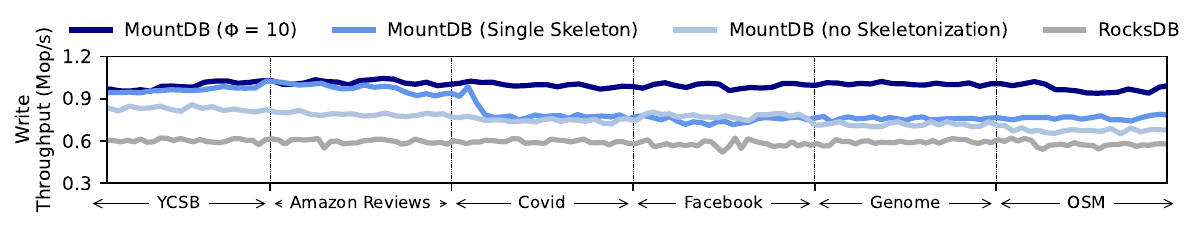} 
    \vspace{-3mm}
    \caption{\textbf{Microbench - Varying Key Distribution.} Write throughput of 1) \mountdb{} ($\phi$=10) - a new skeleton created every 10 flushes, 2) \mountdb{} - single skeleton, 3) \mountdb{} without skeletonization, and 4) \rocksdb{} \skiplist{}. The key distribution changes every $\sim$16.7M operations to cycle through different dataset key distributions. Completion  time: \mountdb{} ($\phi$=10) = 98 sec, \mountdb{} (Single Skeleton) = 127 sec, \mountdb{} (no Skeletonization) = 146 sec, and \rocksdb{} = 161 sec. \shorten{}}
    \label{fig:exp_skeletonization}
\end{figure*}

In this section, we provide a detailed analysis of \mountdb{}'s skeletonization optimization. We study the impact of skeletonization 
under fluctuating key distributions.
This experiment shows a write-only workload that cycles through all six datasets shown in Figure~\ref{fig:cdfs}, issuing approximately 16.7M inserts per dataset for a total of 100M inserts. For each dataset interval, we observe roughly 35 Memtable flushes, during which the systems are subjected to different key distributions. In all settings, 16 workers issue writes concurrently.
We compare four configurations: 1) \mountdb{} ($\phi$=10) uses skeletonization with periodic refresh. In this setting, the \alexol{} index of the Memtable is recreated from scratch based on the data of the previous immutable Memtable every 10 flushes.  All other times, the skeleton is reused. 2) \mountdb{} (Single Skeleton) reuses the same (i.e., first) skeleton after each Memtable flush.  3) \mountdb{} (no Skeletonization), where the learned index is rebuilt from scratch for every Memtable. 4) \rocksdb{}’s \skiplist{}, as a baseline. 

Figure~\ref{fig:exp_skeletonization} shows the measured write throughput as a function of write operations completed. 
\mountdb{} ($\phi$=10) maintains a steady throughput across all six datasets, sustaining around 0.95–1.05 Mops/s throughout the 100M inserts. The periodic skeleton refresh prevents costly full reconstructions, while allowing \mountdb{} to adapt to rapidly shifting key distributions without throughput penalties. 
The single-skeleton variant performs well for the first, simpler distributions (e.g., YCSB and Amazon Reviews), but begins to struggle as the workload transitions to more challenging distributions such as Covid, Facebook, Genome, and OSM. In these phases, throughput drops noticeably, showing that relying on a fixed skeleton limits adaptability to large CDF shifts. The skeleton created when the distribution was simple cannot provide adequate performance for more complex distributions. 
The no Skeletonization configuration exhibits cold-start inefficiencies throughout the experiment, as pointed out in Section~\ref{sec:design_tensions}.
Lastly, \rocksdb{}’s \skiplist{} remains stable and insensitive to changing distributions but maintains a lower steady-state throughput of 0.6–0.7 Mops/s, falling below all other configurations.



\subsection{Read Performance Analysis}
\label{sec:read-performance-eval}

\begin{figure*}[t]
    \centering    
    \includegraphics[width=1.01\textwidth]{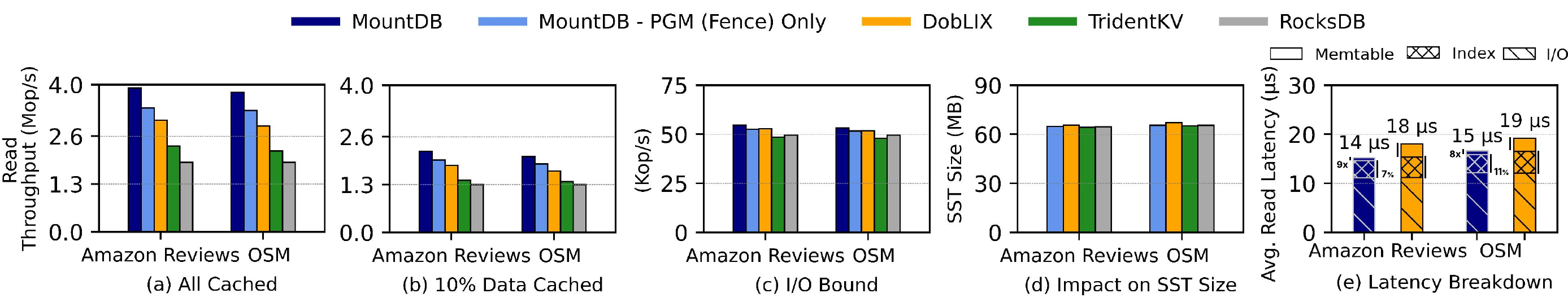} 
    \caption{Read Performance Analysis. Workload with (a) All data cached (fits in memory), (b) 10\% Data cached, (c) I/O-bound with page cache bypassed, (d) Impact on SST size, and (e) Read latency breakdown on a balanced workload showing contributions from in-memory and on-disk index.}
    \label{fig:exp_read_performance_analysis}
\end{figure*}

Next, we  analyze the read path to understand the sources of performance improvements and isolate the contributions of in-memory and on-disk indexing. Figure~\ref{fig:exp_read_performance_analysis} reports read throughput under different caching regimes (a--c), along with the impact on SST size (d) and a latency breakdown of the read path (e). 


We first consider a scenario where the entire working set is cached. In this case, \mountdb{} achieves the highest throughput, outperforming the baselines by approximately 1.34–1.52$\times$ across datasets. Importantly, even the \pgmfence{}-only configuration delivers a considerable improvement, exceeding \doblix{} by 1.16–1.22$\times$ with the gap more pronounced on the harder OSM distribution. In this setting, performance is strongly influenced by index size 
, where \pgmfence{} is roughly an order of magnitude smaller than \doblix{}, resulting in more efficient in-memory lookups and more substantial gains relative to the other baselines.

We next evaluate a partially cached scenario where only 10\% of the working set fits in memory (Figure~\ref{fig:exp_read_performance_analysis}(b)). \mountdb{} maintains consistent advantages, with throughput improvements of roughly 1.2–1.3$\times$ across datasets. Compared to the fully cached scenario, the gap narrows as disk access begins to dominate. For the \pgmfence{}-only configuration, the gains relative to \doblix{} remain in the range of 8–12\%, with larger improvements again observed on the harder distribution. 

To isolate the impact of disk access, we further evaluate an I/O-bound configuration (Figure~\ref{fig:exp_read_performance_analysis}(c)) where each lookup that is not served from the Memtable results in a storage access. To simulate this setting, we enable the $\texttt{use\_direct\_reads}$ flag in $\texttt{db\_bench}$, bypassing the page cache and forcing I/O on every SST access. In this case, throughput across systems becomes virtually the same, as disk latency dominates the overall cost. \mountdb{} still shows slight improvements due to a fraction of lookups being satisfied directly from the Memtable; however, since most requests incur I/O, the overall gains remain modest. 

Figure ~\ref{fig:exp_read_performance_analysis}(d) examines the relative impact of index structures on SST size. Across all systems, the index constitutes only a small fraction of the total SST footprint. Specifically, \tridentkv{} has the smallest relative overhead at approximately 0.025\% (16 KB) of the SST size (64 MB), followed by \mountdb{}’s \pgmfence{} at 0.26\% (172 KB), and \rocksdb{}’s fence pointer table at 0.45\% (296 KB). \doblix{} incurs the largest index overhead, reaching up to 2.42\% (1.55 MB), which is about $\sim9\times$ bigger than \pgmfence{}.

Finally, Figure~\ref{fig:exp_read_performance_analysis}(e) shows a breakdown of the average read latency for the balanced workload presented in Figure~\ref{fig:exp_large_scale_concurrent} (middle row), comparing \mountdb{} and \doblix{}. On average, the latency improvement is primarily driven by faster disk-level lookup, where \pgmfence{} provides approximately 7–11\% faster block localization than \doblix{}. Although Memtable lookup in \mountdb{} is 8–9$\times$ faster than the \skiplist{}, the absolute latency of this stage is small; therefore, its contribution to end-to-end latency is less pronounced once I/O becomes part of the critical path.

\subsection{Overhead of Learned Index Creation}
\label{sec:compaction-eval}

\begin{figure}[t]
    \centering
    \includegraphics[width=1.01\linewidth]{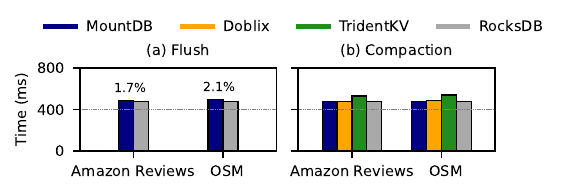} 
    \caption{\textbf{Overhead of Learned Index Techniques on Background Operations.}} 
    \label{fig:exp_overhead}
\end{figure}

Lastly, we evaluate the runtime overhead introduced by learned indexing during background operations. Figure~\ref{fig:exp_overhead} reports the time spent in flush and compaction phases, isolating the cost of index construction. \shorten{}

During a flush, \mountdb{} performs skeletonization to initialize the Memtable (\alexol{}) for future instances. To quantify its cost, we measure flush time with and without this step. As shown in Figure~\ref{fig:exp_overhead}(a), the additional overhead remains small, increasing flush latency by only 1.7\% on Amazon Reviews and 2.1\% on OSM. This demonstrates that preserving structural knowledge across Memtables incurs negligible runtime cost while enabling faster steady-state performance.

During compaction, all systems construct their on-disk index structures. Figure~\ref{fig:exp_overhead}(b) shows that \mountdb{}, \doblix{}, and \rocksdb{} exhibit nearly identical compaction times across both datasets, indicating that building learned index structures introduces no additional overhead. Overall, compaction latency remains within a narrow range across systems, confirming that learned index construction does not meaningfully affect background operations.


\section{Conclusion}
\label{sec:conclusion}
We introduced \mountdb{}, the first LSM-based key-value store to integrate learned indexes across both in-memory and disk levels. By combining \alexol{} with skeletonization for Memtables and \pgmfence{} for SSTs, \mountdb{} improves read and write performance on various real-world data distributions and workloads. \mountdb{} is a step towards motivating learned index adoption in production environments by emphasizing simplicity of integration.




\bibliographystyle{ACM-Reference-Format}
\bibliography{ref}

\end{document}